\documentclass[aps, prd,  twocolumn, preprintnumbers, nofootinbib]{revtex4-1}
\usepackage{myoptions}

\begin{document}

\title{Some results on the shape dependence of entanglement and R\'enyi entropies}

\author{ Andrea Allais}
\affiliation{Department of Physics,\\ Harvard University,\\ Cambridge, MA 02138, USA}
\author{M\'ark Mezei}
\affiliation{Center for Theoretical Physics, \\ Massachusetts Institute of Technology,
Cambridge, MA 02139\\ }

\preprint{MIT-CTP 4569}

\begin{abstract}

\noindent We study how the universal contribution to entanglement entropy in a conformal field theory depends on the entangling region. We show that for a deformed sphere the variation of the universal contribution is quadratic in the deformation amplitude. We generalize these results for R\'enyi entropies. We obtain an explicit expression for the second order variation of entanglement entropy in the case of a deformed circle in a three dimensional CFT with a gravity dual. For the same system, we also consider an elliptic entangling region and determine numerically the entanglement entropy as a function of the aspect ratio of the ellipse. Based on these three-dimensional results and Solodukhin's formula in four dimensions, we conjecture that the sphere minimizes the universal contribution to entanglement entropy in all dimensions.
\end{abstract}

\maketitle

\section{Introduction and summary of results}

In recent years great attention has been devoted to the properties of the entanglement and R\'enyi entropies  of quantum systems in their ground states. In a system with a local Hamiltonian, the Hilbert space is split between the degrees of freedom of a spatial region $V$ and its complement $\overline{V}$, and the entanglement entropy (EE) is defined as the von Neumann entropy of the reduced density matrix of one of the subsystems:
\es{EEDef}{
 S_E = -{\Tr}_V \rho_V \log \rho_V\,, \qquad \rho_V = {\Tr}_{\overline{V}} \ket{\mathrm{gd}}\bra{\mathrm{gd}}\,.
}
The R\'enyi entropy is a generalization of EE defined by
\es{Renyi}{
S_q=-{1\ov q-1}\, \log \Tr \rho_V^q \ .
}
For a normalized density matrix $\Tr \rho_V = 1$, the EE \eqref{EEDef} is obtained in the limit $q\to1$.

By now several properties of these quantities have been uncovered for various classes of systems, and this has led to substantial progress in disparate fields, from numerical methods to the classification of phases of matter. See~\cite{Ryu:2006ef,Amico:2007ag,Calabrese:2009qy,Casini:2009sr} for reviews from different viewpoints.

\begin{figure}
\begin{center}
\includegraphics[scale=0.5]{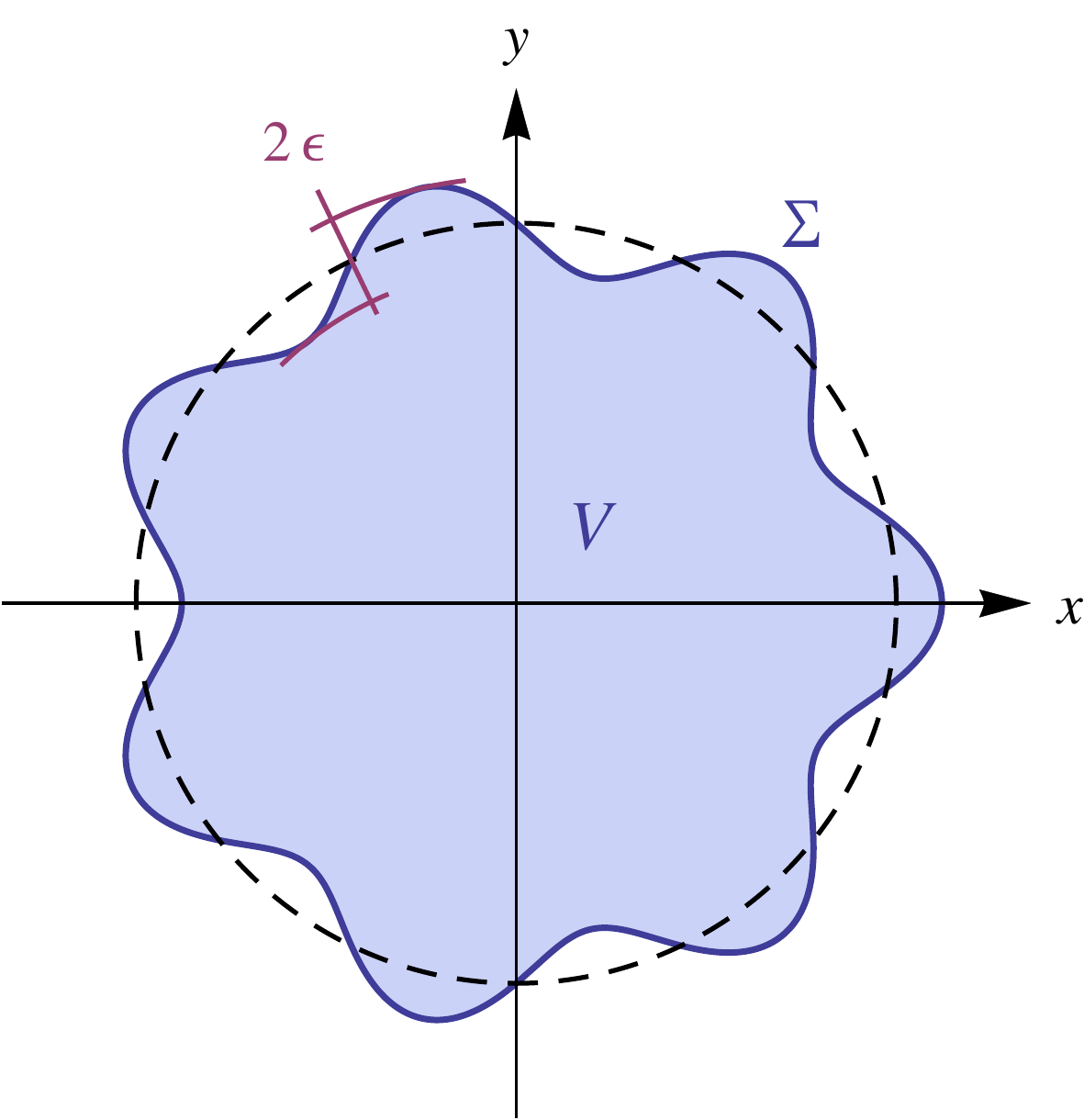}
\end{center}
\caption{\label{fig:wiggly_region}Perturbed circle (\ref{garesult}) as entangling region $V$, $\Sigma = \partial V$. The correction to the universal coefficient $s_{3}$ is of order $\epsilon^2$ and it is given by  \eqref{garesult} in a holographic CFT$_3$.}
\end{figure}

Here we consider the ground state EE of conformal field theories (CFTs), and we focus on their dependence on the shape of the entangling region $V$. More precisely, we consider in 3 and 4 spacetime dimensions the shape dependence of the universal coefficients $s_3$, $s_4$ that appear in the well known expansion
\es{Divergences}{
 &S = \# \frac{R}{\delta} - s_3 + \mathcal O\left(\frac{\delta}{R}\right)\quad \text{for $d = 3$}\,, \\ 
 &S = \# \frac{R^2}{\delta^2} - s_4 \log \frac{R}{\delta} + \mathcal O\left(\frac{\delta^2}{R^2}\right) \quad \text{for $d = 4$}\,,
}
where $\delta$ is a short distance regulator, $R$ is the linear size of the entangling region $V$, and $\#$ stands for non-universal coefficients. 
A similar formula with  $q$-dependent coefficients applies to R\'enyi entropies. 

While in even dimensions the universal coefficient multiplies a logarithmic divergence, and hence its shape dependence is given by a local functional of the geometric invariants of $\Sigma = \partial V$,\footnote{See~(\ref{genes}) for the case of EE in $d=4$ ~\cite{Solodukhin:2008dh}.} in odd dimensions it is a fully non-local functional of the entangling region, whose computation is rather challenging.

Here we show that, for a general CFT in any dimension, the variation of the universal term for a perturbed sphere $r(\Omega) = R_0(1  + \epsilon f(\Omega))$ is second order in the perturbation:
\begin{equation}\label{eq:second_order}
 s_d = s_d^{(0)} + \epsilon^2 s_d^{(2)} + \mathcal{O}(\epsilon^3)\,.
\end{equation} 
Based on ideas of the work~\cite{aitor}, we generalize these results for R\'enyi entropies.

We obtain this second order variation for the EE of a perturbed circle (Fig.~\ref{fig:wiggly_region}) in a $d = 3$ CFT with a gravity dual:
\es{garesult}{
 &r(\theta) = R_0\left[1 + \epsilon \sum_{n} \left(a_n \cos n\theta +b_n \sin n\theta \right)\right]\,, \\
 &\tilde s_3=2\pi\left[1+\ep^2\,\sum_{n}  {n(n^2-1)\ov4} \le(a_n^2+b_n^2\ri)\right]\,,
}
where $\tilde s_d \equiv (4G_N / L^{d-1}) s_d$, with $L$ the radius of the dual AdS$_{d+1}$ and $G_N$ Newton's constant.

\begin{figure}
\begin{center}
\includegraphics[scale=0.5]{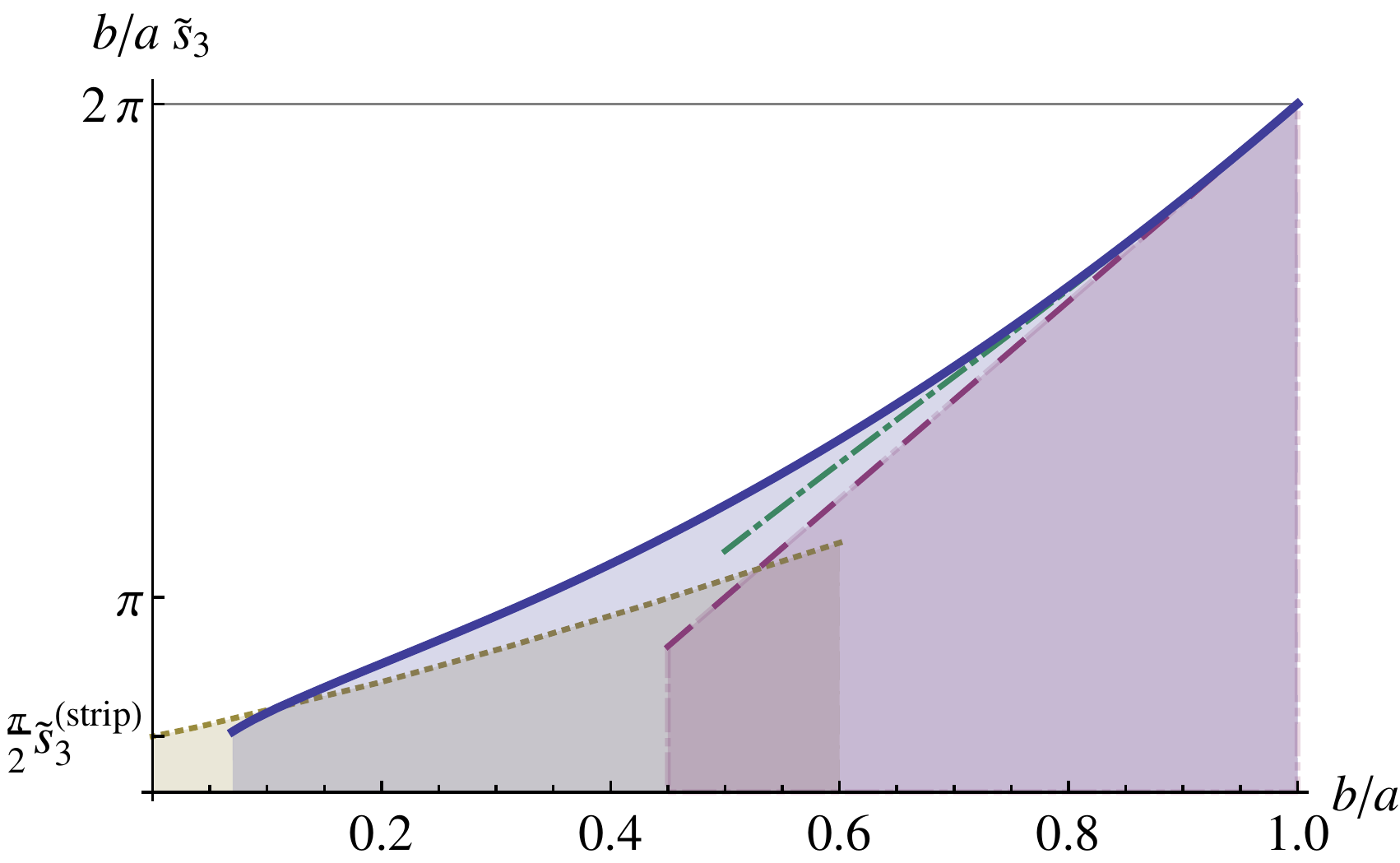}
\end{center}
\caption{\label{fig:ellipse} Universal coefficient $\tilde s_3$ for an elliptic entangling region with semi-axes $a$, $b$, in a $d = 3$ holographic CFT. The blue, solid curve is a tight lower bound obtained numerically. The red, dashed curve $\tilde s_3=2 \pi$ is a lower bound set by the area of an ellipsoid (\ref{eq:ellipsoid_trial}). The yellow, dotted curve is a lower bound set by the area of a deformed strip \eqref{strip}. The green, dash-dot curve $\tilde s_3 = 2 \pi \left[1 + {3\ov8}(1 - b/a)^2\right]$ is an approximation obtained by considering perturbations of a circle~\eqref{gaEllipse}. It is not a bound.}
\end{figure}

We also consider an ellipse of semi-axes $a$, $b$ as entangling region, still in a $d = 3$ CFT with a gravity dual. Fig.~\ref{fig:ellipse} displays various analytic and numerical lower bounds on $\tilde s_3$, as a function of the aspect ratio $b/a$. We find that $\tilde s_3$ smoothly interpolates between the value for a circle and the value for an infintely long strip, as the aspect ratio goes to zero. In particular we have:
\begin{align}\label{eq:ellipse_bounds}
 &\tilde s_3 \geq 2 \pi\,, && \frac{b}{a}\tilde s_3 \geq  \frac{2\pi^2\Gamma\left(\frac34\right)^2}{\Gamma\left(\frac14\right)^2} \equiv \frac{\pi}{2} \tilde s_3^{\mathrm{(strip)}}\,,
\end{align}
where the first bound is saturated at $b/a = 1$ and the second at $b/a = 0$. The very tight lower bound shown in Fig.~\ref{fig:ellipse} with a blue solid line is obtained numerically. It is basically saturated for $b/a \gtrsim 0.1$. From~\eqref{garesult} we can also determine the approach to $b/a = 1$:
\es{gaEllipse}{
\tilde s_3 \sim 2\pi\left[1 + {3\ov8}\left(1 - \frac{b}{a}\right)^2\right] \quad \text{for $b/a \to 1$} .
}

From~\eqref{eq:second_order} it clear that the sphere is a stationary point for the universal term in EE among all shapes. From~\eqref{garesult} we conclude that in holographic theories in $d=3$ it is a local minimum, while the numerical results for an ellipse (see Fig.~\ref{fig:ellipse}) hint at it being a global minimum. 

In $d=4$ the sphere is a global minimum in the universal term for EE~\cite{Astaneh:2014uba}. Let us repeat the analysis here. Solodukhin's formula~\eqref{genes} determines the universal piece for all CFTs~\cite{Ryu:2006ef,Solodukhin:2008dh}:
\es{genes}{
s_4 &= { a_4\ov 180} \int_\Sig d^2 \sig \sqrt{\ga} \, E_2 + {c_4\ov 240\pi} \int_\Sig d^2 \sig \sqrt{\ga} \, I_2\,,\\
 I_2 &=    K_{ab} K^{ab} - \ha K^2  \,,
}
 where $a_4$ and $c_4$ are coefficients of the trace anomaly,\footnote{We normalize $a_4$ and $c_4$ so that they both equal one for a real scalar field.} $E_2$ is the Euler density normalized such that $\int_{S^2} d^2 \sig \sqrt{\ga} \, E_2=2$,  $\ga$ is the induced metric, and $K$ the extrinsic curvature. Because the first term in~\eqref{genes} is topological, shapes continuously connected to $S^2$ give the same contribution. It is easy to see that $I_2$ is nonnegative and vanishes only for the sphere. Thus, we showed that the sphere minimizes the universal term in EE.

The above evidence lead us to conjecture  that, in a CFT, {\it the sphere minimizes the universal contribution to EE in all dimensions} among shapes continuously connected to it.\footnote{We thank Hong Liu for crucial discussions on this topic and Eric Perlmutter for discussions on the topology of $\Sig$.} It is then natural to use the EE across a sphere as a c-function~\cite{Casini:2004bw,Myers:2010tj,Komargodski:2011vj,Casini:2012ei}.
It would be nice to provide more checks for this conjecture; one could investigate e.g.~higher dimensional cases, second order perturbations around in the CFT setup, and whether the sphere still minimizes the universal term to EE away from a CFT fixed point.\footnote{See~\cite{Klebanov:2012yf} for the shape dependence in gapped theories.} We believe these are fascinating topics to explore.

The rest of this paper presents a derivation of these results, organized as follows. In section \ref{sec:perturbed_sphere} we derive (\ref{eq:second_order}) using CFT techniques. In section \ref{sec:perturbed_circle} we derive (\ref{garesult}) and an analogous result for $d = 4$. Section \ref{sec:ellipse} derives the analytic bounds (\ref{eq:ellipse_bounds}) for the elliptic entangling region in a holographic CFT$_3$. Section \ref{sec:generic_surface} describes how to establish tight numerical bounds on $s_3$ for a generic entangling region in a holographic CFT$_3$, and in particular the numerical bound in Fig.~\ref{fig:ellipse}.

{\bf Note added:} During the finalization of this paper, we learned about the work~\cite{aitor}, which will appear soon.

\section{Perturbed sphere in a generic CFT}
\label{sec:perturbed_sphere}

\subsection{First order corrections to entanglement entropy}

In this section, using general conformal field theory arguments we investigate EE of a deformed sphere. The parameter of the deformation will be denoted by $\ep$. We show that the contribution to the universal term in the entropy linear in $\ep$ vanishes.

In polar coordinates
\begin{equation}
 \dd s^2= -\dd t^2+\dd r^2+r^2\dd \Omega^2\,,
\end{equation} 
we take the entangling surface $\Sigma$ to be $r=R \le[1+\epsilon f(\Omega)\ri]$. By changing coordinates we can think about the family of these surfaces as being spheres ($r'=R$), and the field theory living in curved space~\cite{Banerjee:2011mg}:
\begin{align}\label{NewCoord1}
\begin{split}
 \dd s^2& =-\dd t^2+\le(\le[1+\epsilon f(\Omega)\ri]\dd r'+ \ep r' \p_\Omega f(\Omega)\,  \dd \Omega \ri)^2\\
&\phantom{=}\ +r'^2\le[1+\epsilon f(\Omega)\ri]^2\dd \Omega^2\\
&=-\dd t^2+\le(\delta_{ij} +\ep h_{ij}+\sO(\epsilon^2)\ri) \dd x^i \dd x^j \,,
\end{split}
\end{align}
with 
\es{hExplicit}{
h_{ij}\dd x^i \dd x^j &= 2 \le(f \dd r'^2 +  r' \p_\Omega f \, \dd r' \dd \Omega+ r'^2 f \dd \Omega^2\ri) \,,
}
where $i,j$ indices run over spatial directions, while $\mu,\nu$ will run over all spacetime directions. Soon we will introduce a mapping to $\sH=S^1\times \HH^{d-1}$; there we will use $\al,\beta$ as spacetime indices.

From now on we drop the prime from $r'$. An important thing to note is that $h$ is pure gauge:
\es{PureGauge}{
h_{\mu\nu}=2\nabla_{(\mu}\, \xi_{\nu)} \qquad \xi_\nu=(0, r f(\Omega), 0, \dots, 0) \,,
}
where $\nabla_{\mu}$ is the covariant derivative in polar coordinates.

The reduced density matrix in curved space differs from the flat space one. To linear order in the perturbations:
\es{PerturbedEE}{
\rho&=\rho_0+\ep\delta \rho\\
\delta S_E&=-{\ep}\,  {\Tr \le(\delta \rho\, \log \rho_0\ri)} \,,
}
where the subscript $E$ stands for entanglement. To arrive at this formula we used the cyclicity of the trace and the normalization condition $\Tr\rho_0=1$.

The reduced density matrix for a spherical entangling surface is 
given by~\cite{Casini:2011kv}\footnote{An explicit expression for the reduced density matrix is only know in the case of planar and spherical entangling surface. These are the known cases, where there the entanglement Hamiltonian $K_0$ generates a symmetry around $\Sig$.}
\es{EntanglementH}{
\rho_0&=\,\frac{e^{-K_0}}{\Tr e^{-K_0}}\,,\\
K_0&=2\pi \int_{r<R}d\vec{x}\, {R^2-r^2\ov 2R} T^{00}(\vec{x}) \ .
}

\begin{figure}
\begin{center}
\includegraphics[scale=0.5]{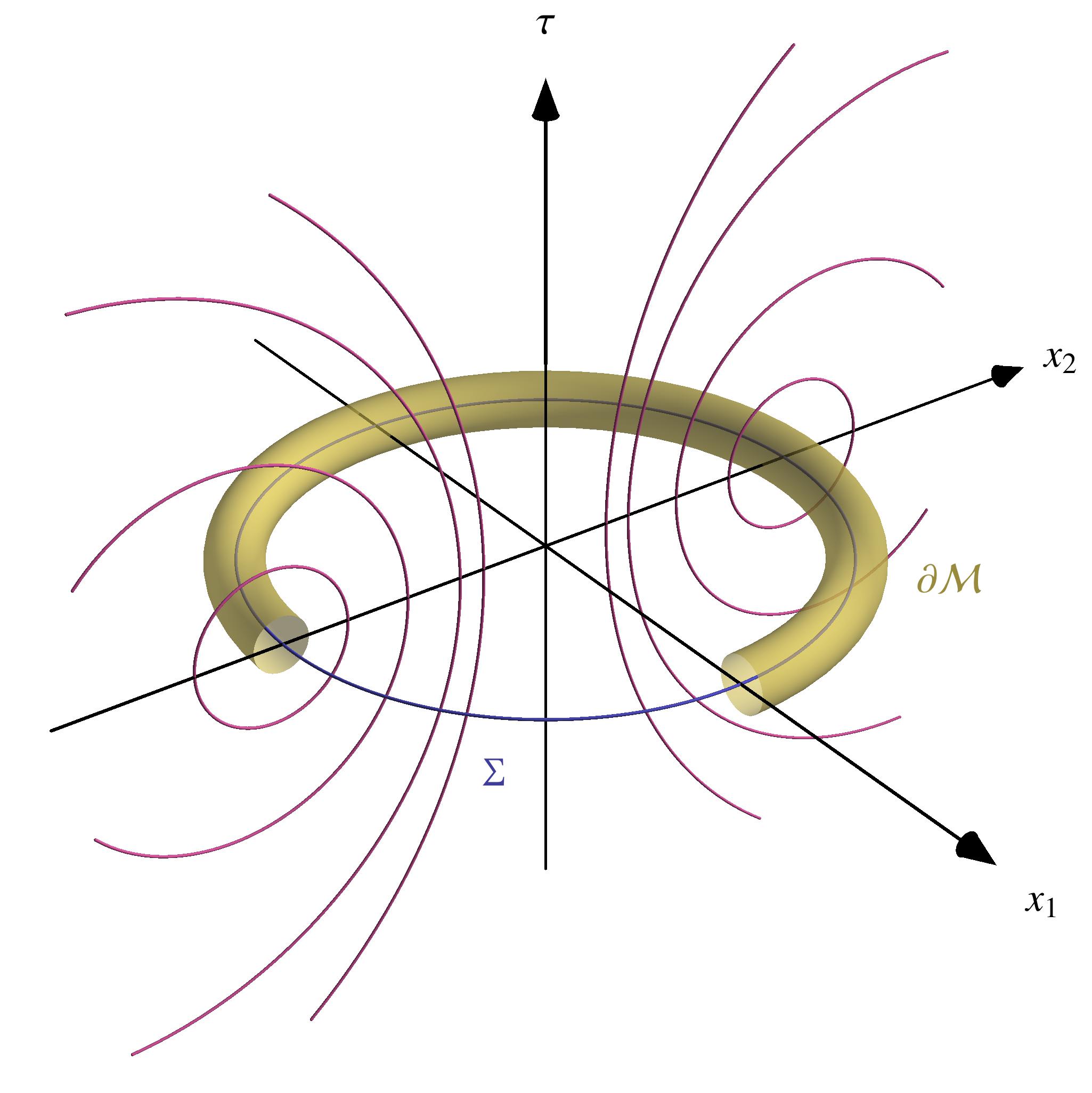}
\end{center}
\caption{\label{fig:MBdy} Geometry of the manifold $\sM$ of~\eqref{deltaRho}. Lines of constant $(u,\Omega)$ are drawn in purple. The entangling surface $\Sig$ is marked by a blue line, and sits at $u=\infty$. We use a regularization procedure with a cutoff $\delta$ that cuts out  a tube centered around $\Sig$ from $\sM$. $\p\sM$ is at constant $u=u_m$, and it has topology $S^1\times S^{d-2}$. It is drawn in yellow. When we map to hyperbolic space the Hamiltonian generates time evolution along the purple lines. $\p\sM$ maps to the boundary of hyperbolic space.}
\end{figure}

Recently,~\cite{Rosenhaus:2014woa} gave an elegant formula for $\de \rho$:
\es{deltaRho}{
\rho_0^{-1}\, \delta \rho&={1 \ov 2 } \int_\sM  h_{\mu\nu} \, \le(T_\sI^{\mu\nu}-\Tr \le(T^{\mu\nu} \rho_0\ri) \ri) \,, \\
T_\sI^{\mu\nu}(v,u,\Om)&\equiv\rho_0^{v/2\pi}  \, T^{\mu\nu}(v,u,\Om) \, \rho_0^{-v/2\pi} \,,
}
where $\sM$ is (Euclidean) $\R^d$, with a tube of size $\delta$ around the entangling surface $\Sigma$ cut out (see fig.~\ref{fig:MBdy}). The removal of this tube serves as a short distance regulator. We like to think of $T_\sI^{\al\beta}(v,u,\Om)$ as the analog of an operator in the interaction picture in weak coupling perturbation theory. Because instead of a time-ordered exponential $\rho_0=e^{-K_0}/\Tr e^{-K_0}$, its fractional powers indeed generate the appropriate time evolution.

Plugging into~\eqref{PerturbedEE} and using the cyclicity of the trace we get:
\es{PerturbedEE2}{
\delta S_E&={\ep\ov 2}\, \int_\sM h_{\mu\nu}\le[\Tr \le(T^{\mu\nu} K_0\, \rho_0\ri)-\Tr \le(T^{\mu\nu}\rho_0\ri)\Tr\le(K_0\,\rho_0\ri)\ri] \,.
}
To lighten the notation we introduce the ``connected'' trace:
\es{ConnectedTrace}{
\Tr \le(T^{\mu\nu} K_0\, \rho_0\ri)_c\equiv\Tr \le(T^{\mu\nu} K_0\, \rho_0\ri)-\Tr \le(T^{\mu\nu}\rho_0\ri)\Tr\le(K_0\,\rho_0\ri) \,.
}
\eqref{PerturbedEE2} can then be written as
\es{PerturbedRenyiLimit}{
\delta S_E&={\ep \ov 2}\, \int_\sM h_{\mu\nu}\,\Tr \le(T^{\mu\nu} K_0\, \rho_0\ri)_c \,.
}
where we used $\Tr\rho_0=1$.\footnote{In $\sM$ the ''connectedness'' of the correlator does not matter, as one point functions in the ground state vanish, hence $\Tr \le(T^{\mu\nu}\, \rho_0\ri)=0$. Nevertheless we kept the disconnected terms to facilitate the transformation to $\sH$, where in even $d$ the transformation rule of these terms cancel the anomalous term coming from the transformation rule of the stress tensor~\cite{Hung:2011nu,Perlmutter:2013gua,Rosenhaus:2014woa}.}

Now we want to make use of the fact that $h$ is pure gauge. We partially integrate to get:
\begin{equation}\label{PartInt}
\begin{split}
 \int_\sM \nabla_{\mu}\, \xi_{\nu}\,  \Tr\le(T^{\mu\nu} K_0\, \rho_0\ri)&=\int_{\p\sM} n_{\mu}\xi_{\nu}\, \Tr\le(T^{\mu\nu} K_0\, \rho_0\ri)\\
&-\int_\sM  \xi_{\nu}\,   \Tr\le(\le(\nabla_{\mu}\,T^{\mu\nu}\ri)K_0\, \rho_0\ri) \,,
\end{split}
\end{equation} 
where $\p \sM$ is shown on Fig.~\ref{fig:MBdy}.
The Ward identity from the conservation of the energy momentum tensor is:
\es{Ward}{
\Tr\le(\le(\nabla_{\mu}\,T^{\mu\nu}\ri)K_0\, \rho_0\ri)&=0\,,
}
hence only the boundary term remains on the right-hand side of~\eqref{PartInt}.\footnote{We have to show that the counter term contributions vanish. From~\eqref{EntanglementH} we see that $K_0$ is an integral of $T^{00}$, and the one point function of $T^{00}$ vanishes by conformal invariance, and~\eqref{Ward} follows.} 
Finally, we are left with
\es{PT4}{
\delta S_E&=\ep\, \int_{\p\sM} n_{\mu}\xi_{\nu}\,\Tr \le(T^{\mu\nu}\, \rho_0^{q}\ri)_c \,.
}
The non-universal contributions to the entropy come from the entanglement of degrees of freedom on the cutoff scale $\delta$. According to~\eqref{PT4} the only contribution to $\delta S_E$ comes from the cutoff-size region $\p\sM$, hence we already anticipate that $\delta S_E$ will not contain a universal piece. In the following, we confirm this intuition by explicit calculation.

We make a cautionary remark about boundary contributions here. We have not been careful about imposing boundary conditions on $\p\sM$. In the context to mapping to hyperbolic space, it is known that these boundary terms contribute to the thermal entropy~\cite{Casini:2011kv}. See~\cite{Lewkowycz:2013laa} for additional discussions.  We leave the analysis of this subtlety to future work.

\subsection{Calculation in hyperbolic space}

The calculation of $\delta S_E$ is most easily done by going to hyperbolic space, $\sH=S^1\times \HH^{d-1}$. We work in Euclidean signature, and set the radius of hyperbolic space, or equivalently of the entangling surface to 1.
The change of coordinates
\es{MapToH}{
\tau&={\sin (v)\ov \cosh u + \cos(v)}\,,\\
r&={\sinh u \ov \cosh u + \cos(v)}\,,
}
leads to the unperturbed metric 
\es{ConfEqMetric}{
ds^2_0&=\omega^2\le[dv^2+du^2+\sinh^2u \,d\Omega_{d-2}^2\ri]\\
\omega&={1\ov \cosh u + \cos(v)} \,.
}
The range of the coordinates are $u\in[0,\infty),\, v\in[0,2\pi)$.
We can get rid of the conformal factor $\omega^2$ by a Weyl scaling, and the remaining line element is $\sH$. Actually, there is a conformal transformation relating the operators on $\sM$ to those on $\sH$, and implementing this transformation on the entanglement Hamiltonian $K_0$ we obtain the Hamitonian on $\HH^{d-1}$, $2\pi H$~\cite{Casini:2011kv}. Going through these steps we obtain~\eqref{PerturbedEE2} in $\sH$~\cite{Rosenhaus:2014woa}:
\es{PT5}{
\delta S_E&={\ep\ov 2}\, \int_{\p\sH} \omega^{-2}\, h_{\al\beta}\, \Tr \le[T^{\al\beta}\le(2\pi H\, e^{-2\pi q H}\ri)\ri]_c \,.
}
where we used the conformal transformation rule of the stress tensor.\footnote{We emphasize that $h_{\al\beta}$ is what we get by the coordinate transformation~\eqref{MapToH} and does not change under Weyl scaling. Alternatively, we could also use that $\delta S_E$ should be given by a Weyl invariant expression, and under a Weyl scaling the metric deformation also changes. In the latter way of thinking the $\om^{-2}$ factor comes from the transformation of $h$.} 

We can make the next step in two equivalent ways; we can either perform a conformal transformation on~\eqref{PT4} or we can use that $h$ is pure gauge,\footnote{Note that~\eqref{PureGauge} only holds in flat space, in the conformally related $\sH$ there are additional terms (due to the change of the covariant derivative under Weyl scalings). They conspire to yield and integrand which is again a total divergence.} and integrate partially to obtain
\es{PT6}{
\delta S_E&=\ep\, \int_{\p\sH} \omega^{-2}\, n_{\al}\xi_{\beta}\, \Tr \le[T^{\al\beta}\le(2\pi H\, e^{-2\pi q H}\ri)\ri]_c\, ,
}
where $\p \sH$ is $S^1\times S^{d-2}$ at constant radial coordinate $u=u_m$.

The integrals can now be evaluated. From the tracelessness and conservation of the stress tensor it follows that $\Tr \le[T^{\al}_{\,\,\beta}\le(2\pi H\, e^{-2\pi q H}\ri)\ri]_c$ is position independent, and the non-zero elements are~\cite{Rosenhaus:2014woa,Perlmutter:2013gua}:
\es{TTCorrelator2}{
\Tr \le[T^{v}_{\,\,v}\le(2\pi H\, e^{-2\pi H}\ri)\ri]&=-{(d-1)\om_{d+2}\ov 2^{d+1} \pi \, d}\, C_T\\
\Tr \le[T^{I}_{\,\,J}\le(2\pi H\, e^{-2\pi H}\ri)\ri]&={\om_{d+2}\ov 2^{d+1} \pi \, d}\, C_T \, \delta^I_J \,,
}
where $I,\, J$ run over $\HH^{d-1}$ and $C_T$ is the coefficient of the stress tensor two-point function:
\es{TT2Point}{
\<\, T_{\mu\nu}(x)\, T_{\rho\lambda}(0)\>_{\R^d}&={C_T \ov x^{2d}}\,\le[\frac12 \le(I_{\mu\rho} I_{\nu\lambda}+I_{\mu\lam} I_{\nu\rho}\ri)-{\delta_{\mu\nu} \delta_{\rho\lambda} \ov d}\ri]\,, \\
I_{\mu\nu}&\equiv \delta_{\mu\nu} -2 {x_\mu x_\nu\ov x^2} \,,
}
and
\es{omd}{
 \om_d={2\pi^{(d+1)/2}\ov\Gamma\le(d+1\ov2\ri)}
 }
 is the volume of $S^d$.

In $\sH$ coordinates $n=(0,1,0,\dots)$ and $\xi=f\, \omega^{3}\sinh u$ $\le(\sin v \sinh u,\, 1+\cos v \cosh u,\, 0, \dots\ri)$. Plugging these formulae in~\eqref{PT6}, and not forgetting about the volume element on $\p \sH$ that we omitted in the above formulae, we get:
\begin{equation}\label{BdyTerms2}
\begin{split}
\delta S_E=&{\ep\,  \om_{d+2}\ov 2^{d+1} \pi\, d}\, C_T\,\,  \le[\int_{S^{d-2}}  d\Omega_{d-2}\ f \ri]\\\times &\int_{S^1}dv \ \om(v,u_m)\, (1+\cos v \cosh u_m) \, \sinh^{d-1}u_m \\
=&{\ep\,  \om_{d+2}\ov 2^{d+1}\, d}\, C_T\,  \le(2 e^{-u_m}\sinh^{d-1}u_m \ri)  \le[\int_{S^{d-2}}  d\Omega_{d-2}\ f \ri] \,. 
\end{split}
\end{equation} 
It can be checked  that the original expression~\eqref{PT5} evaluates to the same answer, if we plug in the explicit form of $h$.

\subsection{Generalization to R\'enyi entropies}

The above discussion can be generalized to the case of R\'enyi entropies. The upcoming paper~\cite{aitor} develops perturbation theory for R\'enyi entropies well beyond what we consider here, and its authors suggested to us to generalize the EE results to R\'enyi entropies. The formulae below have some overlap with~\cite{aitor}, but were obtained independently.

The change in the reduced density matrix~\eqref{PerturbedEE} induces a change in the R\'enyi entropies. To linear order we get:
\es{PerturbedRenyi}{
\delta S_q&=-{\ep q\ov q-1}\,  {\Tr \le(\delta \rho\, \rho_0^{q-1}\ri)\ov \Tr\rho_0^q} \ .
}
The $q\to1$ limit of this formula gives~\eqref{PerturbedEE}. As we have a formula for the operator $\delta\rho$~\eqref{deltaRho}, we can follow the same steps as in previous subsections to arrive at
\es{PT7}{
\delta S_q&=-{\ep q\ov (q-1)\Tr\rho_0^q}\, \int_{\p\sH} \omega^{-2}\, n_{\al}\xi_{\beta}\, \Tr \le(T^{\al\beta}\, e^{-2\pi q H}\ri)_c\,.
}

The traces that we need can be argued to have the same form as~\eqref{TTCorrelator2}, except that the overall constant is not known:
\es{TTCorrelator3}{
\Tr \le(T^{v}_{\,\,v}\, e^{-2\pi q H}\ri)_c&=-(d-1)\al_q\,,\\
\Tr \le(T^I_{\,\,J}\,e^{-2\pi q H}\ri)_c&=\al_q\, \delta^I_J \,,
}
where $\al_q$ is a $q$-dependent undetermined constant. Finally, we obtain
\begin{equation}\label{BdyTermsRenyi}
\begin{split}
\delta S_q=&-{\ep  q\, \al_q  \pi \ov (q-1)\Tr\rho_0^q}\, \le(2 e^{-u_m}\sinh^{d-1}u_m \ri) \\
\times& \le[\int_{S^{d-2}}  d\Omega_{d-2}\ f \ri] \,. 
\end{split}
\end{equation} 
It can be checked  that the $q\to1$ limit of these expressions gives back the EE results.

\subsection{Analysis of the results}

Let us analyze the result. $\int_{S^{d-2}}  d\Omega_{d-2}\ f $ picks out the constant piece from the spherical harmonic decomposition of $f(\Om)$, which is just a change of radius $R$. Changing the radius does not result in the change of the universal piece in a CFT, however it changes the divergent pieces.\footnote{See~\eqref{Divergences} for the divergence structure and the universal pieces in $d=3,4$. } We conclude that $\delta S_q$ {\it does not contain a universal piece}.

To see this more explicitly from~\eqref{BdyTerms2}, we have to express $u_m$ in terms of the field theory cutoff $\delta$. We do not know what the exact relation between these two quantities is, only
its leading behavior
\es{CutoffRelation}{
{\delta\ov 2R}=e^{-u_m}+\dots \  ,
} 
where we have reintroduced the radius of $\mathbb{H}^{d-1}$, or equivalently the size of the entangling region.
The  relation~\eqref{CutoffRelation} can be motivated from the from the coordinate transformation~\eqref{MapToH} by setting $\tau=0$, going to $\delta$ distance to the entangling region at $r=R-\delta$, and relating this to the boundary of $\mathbb{H}^{d-1}$, See Fig.~\ref{fig:MBdy}. However, we should not take this argument literally, as it would give the relation~\cite{Casini:2011kv}
\es{AltCutoffRelation}{
{\delta\ov R}=1-{\sinh u_m\ov \cosh u_m+1} \  .
}
Although this expression gives the same leading behavior as~\eqref{CutoffRelation}, it contains all both even and odd powers of $e^{-u_m}$. However, even powers would result in the change of universal terms through~\eqref{BdyTerms2}, hence they are not allowed.\footnote{Even powers of $e^{-u_m}$ in~\eqref{CutoffRelation} would also invalidate the results of~\cite{Klebanov:2011uf}.} 
We conclude that~\eqref{CutoffRelation} can only involve odd powers of $e^{-u_m}$, so going through the explicit analysis of~\eqref{BdyTerms2} we learned something about~\eqref{CutoffRelation}.

 In the regularization scheme we defined by cutting out a tube around $\Sig$, we obtain
\es{AreaCorrection}{
\delta S\propto&\,{\ep}f_0\, \le[\le(R\ov \delta\ri)^{d-2}+\#\le(R\ov \delta\ri)^{d-4}+\dots\ri] \,,
}
where there the powers of $R/\delta$ decrease in steps of 2, and hence no universal terms occur. We introduced $f_0={1\ov \om_{d-2}}\,\int_{S^{d-2}}  d\Omega_{d-2}\ f $ for the constant mode of $f$. 

In the EE case, we determined the prefactor as well:
\es{AreaCorrectionEE}{
\delta S_E=&{\ep}f_0\,{\om_{d+2}\om_{d-2}\ov 2^{d+1} \, d}\, C_T\, \le[\le(R\ov \delta\ri)^{d-2}+\#\le(R\ov \delta\ri)^{d-4}+\dots\ri] \,,
}
Our calculation then makes it possible to determine the coefficient of the area law term as follows. We obtained that changing the radius $R\to R(1+\ep f_0)$ introduces a change in the EE~\eqref{AreaCorrectionEE}.
 We can then reconstruct the coefficient of the area law term
\es{AreaLaw}{
S&_E={\om_{d+2}\om_{d-2}\ov 2^{d+1} \, d(d-2)}\, C_T\, \le(R\ov \delta\ri)^{d-2}+\dots \ .
}
Of course, this result only applies in the particular regularization scheme that we used in this calculation. 

We repeat that we have not been careful about boundary conditions in this calculation. They could potentially give additional contributions to the result~\eqref{AreaLaw}.

In summary, we found that to linear order in the deformation parameter $\ep$ there is no change in the universal term in the entropies. The only change is in the divergent terms, and they are all proportional to the spherical average of the deformation. In particular the entropies do not change, if this average vanishes. In the next section we calculate the $\sO(\ep^2)$ pieces for EE in the holographic setup.

\section{Perturbed circle in a holographic CFT}
\label{sec:perturbed_circle}

According to the Ryu-Takayanagi formula~\cite{Ryu:2006bv,Lewkowycz:2013nqa}, in a QFT with a gravity dual, the EE of a region $\Sigma$ is proportional to the area of the minimal surface in the dual geometry that is homologous to $\Sigma$. For a CFT$_3$ in its ground state, the dual geometry is AdS$_4$:
\begin{equation}\label{eq:AdS_coordinates}
 g = \frac{L^2}{z^2} \left(-\dd t^2 + \dd z^2 + \dd r^2 + r^2 \dd \theta^2\right)\,.
\end{equation} 

We take the entangling region to be a perturbed circle $r = R(\theta)$ with 
\begin{align}\label{eq:perturbed_circle1}
 R(\theta)=R_0+\ep \, A \, \cos\le(n\theta\ri)\,,
\end{align}
and parametrize the surface inside AdS as 
\begin{align}
 r = R(\theta, z)\,, && R(\theta, 0) = R(\theta)\,.
\end{align} 

\begin{figure}[ht]
\begin{center}
\includegraphics[scale=0.5]{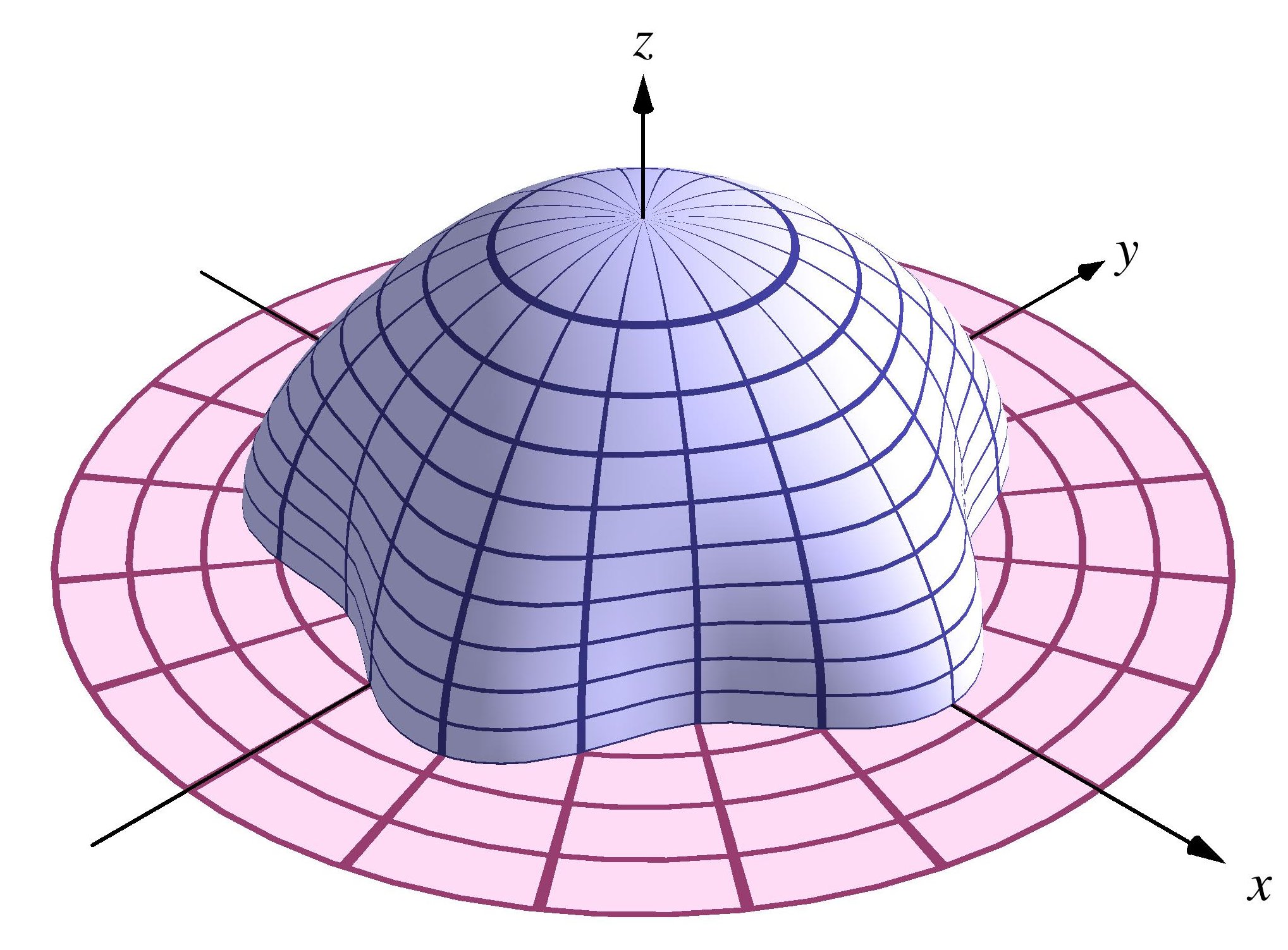}
\end{center}
\caption{\label{fig:perturbed_circle_surface} Actual minimal surface for a perturbed circle (\ref{eq:perturbed_circle1}) with $n = 5$ and $A/R_0 = 0.12$, obtained numerically.}
\end{figure}

We organize the perturbation theory in $\ep$ so that the tip of the minimal surface is at $z=1$ to all orders. Then both $R_0$ and $A$ are non-trivial series in $\ep$, which we evaluate to the order needed to obtain the leading correction to $s_3$:
\begin{align}
R_0=1+{\ep^2\ov 4}+\mathcal{O}(\epsilon^4)\,, &&
A=1+\mathcal{O}(\epsilon) \,.
\end{align}

The minimal surface to $O(\ep^2)$ is:
\es{}{
R(z,\theta)&=\sqrt{1-z^2}\big[1+\ep\, R_1(z)\,\cos\le(n\theta\ri)\\
&+\ep^2\,\le(R_{20}(z)+R_{22}(z)\,\cos\le(2n\theta\ri)\ri)+\dots\big]\\
R_1(z)&=\le(1-z\ov 1+z\ri)^{n/2}\, {1+nz\ov 1-z^2}\\
R_{20}(z)&=\le(1-z\ov 1+z\ri)^{n}\frac{1}{4(1-z^2)^2}\big[1 + 2 n z  \\
& + (3 n^2-2) z^2  + 2n (n^2-1) z^3\big] \,. 
}
(To obtain the leading result we do not need the explicit form of $R_{22}$). See~ \cite{Hubeny:2012ry} for the same solution in different coordinates. Plugging into the area functional we obtain: 
\be
{4G_N\ov L^{d-1}}\, S={2\pi\le(1+\ep^2\, {n^2+1\ov4}\ri)\ov \delta}-2\pi\le(1+\ep^2\, {n(n^2-1)\ov4}\ri) \,,
\ee
where $L$ is the radius of AdS$_4$ and $G_N$ is Newton's constant.
The first term is just the area law term; the length of the entangling region appears expanded to $O(\ep^2)$. From the constant term we read off:
\es{gaResult}{
\tilde s_3&=2\pi\le(1+\ep^2\, {n(n^2-1)\ov4}\ri)\,,\\
\tilde s_3&\equiv {4G_N\ov L^{d-1}}\, s_3\,.
}

\begin{figure}[ht]
\begin{center}
\includegraphics[scale=0.5]{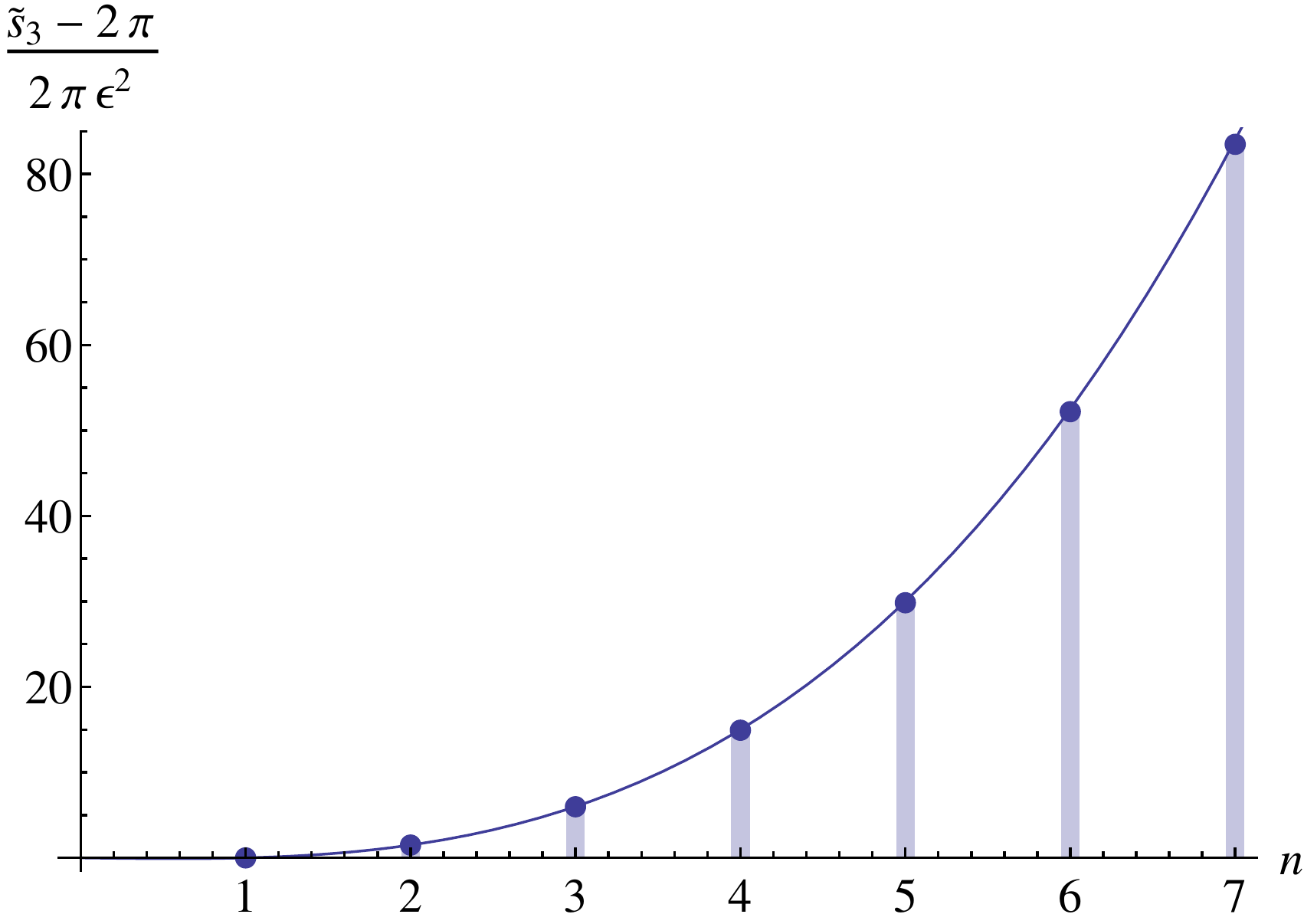}
\end{center}
\caption{\label{fig:wiggly_plot} Universal contribution to the EE for a perturbed circle (\ref{eq:perturbed_circle1}) in a $d = 3$ holographic CFT$_3$. The blue dots are a tight lower bound obtained numerically, the blue line is the analytic result (\ref{gaResult}).}
\end{figure}

As shown in Fig.~\ref{fig:wiggly_plot}, we verified this result numerically, using the methods outlined in section \ref{sec:generic_surface}, and found excellent agreement.

We can easily convince ourselves that for a generic perturbation of the form
\es{GenericPert}{
r&=R_0\le(1+\ep f(\theta)\ri) \,,\\
f(\theta)&=\sum_{n} \le[a_n \cos(n\theta)+b_n \sin(n\theta) \ri]\,,
}
the result is a sum of contributions from different Fourier components:
\es{FourierSum}{
\tilde s_3&=2\pi\le(1+\ep^2\,\sum_{n}  {n(n^2-1)\ov4} \le(a_n^2+b_n^2\ri)\ri) \,.
}
Of course at higher orders in $\ep$ different harmonics mix, and hence the result would no longer be a sum of their individual contributions.

It would be very interesting to perform the CFT calculation of section \ref{sec:perturbed_sphere} to second order in $\ep^2$. It would reveal to what extent~\eqref{FourierSum} is universal. We leave this problem to future work.

For completeness, let us derive the same result for a perturbed sphere in $d = 4$. We can either carry out a perturbative holographic computation, or directly use the result~\eqref{genes}.
Considering an entangling region in the form of a perturbed sphere 
\be
R(\theta)=R\le[1+\ep \, f\le(\theta\ri) \ri]\,,
\ee
we obtain
\es{4dResult}{
s_4&={a_4\ov 90}+ \ep^2 \, c_4\,  g[f]\,,\\
g[f]&\equiv{1\ov 240}\int_0^\pi d\theta\ \sin\theta \le[f''\le(\theta\ri) -\cot \theta\,  f'\le(\theta\ri)\ri]^2 \,.
}
If we further specialize to the case $f_n\le(\theta\ri)=\cos\le(n\theta\ri)$, then $g[f_n]$ can be calculated explicitly
\begin{equation}\label{4dResultCos}
\begin{split}
g[f_n]&={n^2\ov 480}\Bigg[\psi\le(n+\ha\ri)+\psi\le(-n+\ha\ri)\\
&+2\le(\ga+\log 4\ri)+4n^2{2n^2-5\ov 4n^2-1}\Bigg] \,,
\end{split}
\end{equation} 
where $\psi(x)$ is the digamma function. We list the first few values of $g(f_n)$ in Table~\ref{tab:gValues}. As in $d=3$~\eqref{FourierSum}, we can verify that  the quadratic functional~\eqref{4dResult} does not mix different harmonics. A crucial difference from~\eqref{gaResult} is that~\eqref{4dResultCos} is an even function of $n$. The reason for this is that~\eqref{genes},~\eqref{4dResult} are local functionals, while $s_3$ is expected to be a nonlocal functional of $f(\theta)$.

\renewcommand\arraystretch{2}
\renewcommand\tabcolsep{7pt} 
\begin{table}[!h]
\begin{center}
\begin{tabular}{|c||c|c|c|c|c|c|}
\hline
n & $1$ & $2$ &$3$ &$4$ &$5$& $6$\\
\hline
\hline
$g[f_n]$& $ 0$ & $32\ov45$ & $ 128\ov35$ & $  512\ov45$ & $ 56960\ov2079$&$ 844384\ov15015$\\
\hline
\end{tabular}
\end{center}
\caption{The values of $g[f_n]$.}\label{tab:gValues}
\end{table}%

\section{Ellipse in a holographic CFT$_3$}
\label{sec:ellipse}

In this section we derive two analytic lower bounds on the universal coefficient $\tilde s_3$ for an elliptic entangling region in a holographic CFT$_3$. Similarly to the perturbed circle case, it would be interesting to also carry out the computation in a general CFT, or at least in free theories.

Still with reference to the coordinates (\ref{eq:AdS_coordinates}), the entangling region is
\begin{equation}\label{eq:ellipse_R}
 r = R(\theta) \equiv \left(\frac{\cos^2\theta}{a^2}+\frac{\sin^2\theta}{b^2}\right)^{-\frac{1}{2}}\,,
\end{equation} 
where $a$, $b$ are the semi-axes of the ellipse. A trial surface that satisfies the boundary conditions is the squashed hemisphere
\begin{equation}\label{eq:ellipsoid_trial}
z(r,\theta)=a \sqrt{1-\frac{r^2}{R^2(\theta)}}\,.
\end{equation} 
Plugging this expression into the area functional we obtain a bound on $\tilde s_3$, which turns out not to depend on the aspect ratio $b/a$ of the ellipse
\be\label{squashed}
\tilde s_3 \geq 2\pi \,.
\ee
It is well known that the surface (\ref{eq:ellipsoid_trial}) is minimal in the case of a circle ($a = b$), and hence the bound is saturated at this point.

To tackle the opposite limit, $b/a\ll1$, of a very thin ellipse we start from the minimal surface determining the EE of an infinite strip:
\es{}{
y(z)&=-z_t\int_1^{z/z_t}du\, {u^2\ov \sqrt{1-u^4}}\\
z_t&={\Ga\le(\frac14\ri)\ov \sqrt{\pi}\, \Ga\le(\frac34\ri)}\, {l\ov 2} \,.
}
From this we construct the following trial surface satisfying the boundary conditions:
\es{eq:deformed_strip}{
y(z,x)&=-z_t(x)\int_1^{z/z_t(x)}du\, {u^2\ov \sqrt{1-u^4}}\\
&\equiv z_t(x)\, Y\le(z\ov z_t(x)\ri)\\
z_t(x)&={\Ga\le(\frac14\ri)\ov \sqrt{\pi}\, \Ga\le(\frac34\ri)}\, b {\sqrt{1-x^2 / a^2}} \,.
}
By plugging into the area functional
\be
A=2\int_{-a}^a dx\, \int_{\delta}^{z_t(x)} dz\, {1\ov z^2}\sqrt{1+(\p_x y)^2 +(\p_z y)^2}
\ee
we can obtain a bound on $\tilde s_3$. (We have to keep in mind that we only integrate for $x$'s for which $z_t(x)>\delta$. We did not display this to avoid clutter.) Let us use that the $x$ dependence only appears through $z_t(x)$; we rescale $z\equiv z_t \, Z$ to obtain:
\begin{equation}
\begin{split}
A=&2\int_{-a}^a dx\, \int_{\delta/z_t(x)}^{1} dZ\, {1\ov z_t(x) Z^2}\times\\
&\times\sqrt{{1\ov 1-Z^4}+z_t'(x)^2\le(Y(Z)-{Z^3\ov \sqrt{1-Z^4}}\ri)^2 } \end{split}
\end{equation} 
By subtracting the diverging piece of the integrand and adding it back we can get an expression for $\tilde s_3$
\be
\begin{split}
A=&2\int_{-a}^a dx\, \int_{\delta/z_t(x)}^{1} dZ\,\Bigg[ {1\ov z_t(x) Z^2}\times\\
&\times\sqrt{{1\ov 1-Z^4}+z_t'(x)^2\le(Y(Z)-{Z^3\ov \sqrt{1-Z^4}}\ri)^2 }\\
&-{1+Z^2\ov Z^2}\,{ \sqrt{\pi}\, \Ga\le(\frac34\ri)\ov \Ga\le(\frac14\ri) }\,{\sqrt{a^2(a^2-x^2) + b^2 x^2}\ov b(a^2-x^2) }\Bigg]\\
&+{4aE(1-b^2/a^2)\ov \delta}-{ 8\sqrt{\pi}\, \Ga\le(\frac34\ri)\ov \Ga\le(\frac14\ri) } \,,
\end{split}
\ee
where $4aE(1-b^2/a^2)$ is the perimeter of the ellipse,\footnote{$E(x)$ is complete elliptic integral of the second kind.} and thus the divergent term reproduces the area law.
Note that the minimal subtraction $\propto 1/Z^2$ would not have cancelled all divergences. By setting $\delta\to 0$ in the integral we only introduce an $O(\delta)$ error, hence:
\begin{equation}
 \label{strip}
\begin{split}
\tilde s_3\geq&{ 8\sqrt{\pi}\, \Ga\le(\frac34\ri)\ov \Ga\le(\frac14\ri) }-2\int_{-a}^a dx\, \int_{0}^{1} dZ\,\Bigg[ {1\ov z_t(x) Z^2}\times\\ &\times\sqrt{{1\ov 1-Z^4}+z_t'(x)^2\le(Y(Z)-{Z^3\ov \sqrt{1-Z^4}}\ri)^2 }\\
&-{1+Z^2\ov Z^2}\,{ \sqrt{\pi}\, \Ga\le(\frac34\ri)\ov \Ga\le(\frac14\ri) }\,{\sqrt{a^2(a^2-x^2) + b^2 x^2}\ov b(a^2-x^2) }\Bigg]\,.
\end{split}
\end{equation}
Now we can do the logarithmically divergent $x$ integral first. The coefficient of the logarithmic divergent term vanishes when integrated over $Z$. Then we are left with the finite $Z$ integral that depends on $b/a$. We do not know how to evaluate this integral. However, we can calculate the limit
\be\label{striplimit}
\frac{b}{a}\, \tilde s_3\Big\vert_{e\to 1} ={\pi\ov 2}\, \tilde s_3^{\mathrm{(strip)}} \qquad \tilde s_3^{\mathrm{(strip)}} ={4\pi\, \Ga\le(\frac34\ri)^2\ov \Ga\le(\frac14\ri)^2}
\ee 
by expanding the integrand around $b/a= 0$.

Intuitively, this result comes from decomposing the surface into strips of different lengths and adding up their contributions. Because  the EE of a strip is proportional to its length and inversely proportional to its width we get the answer by integrating $\tilde s_3^{\mathrm{(strip)}}{dx/ 2y(x)}$:
\be
\tilde s_3\to \tilde s_3^{\mathrm{(strip)}}\int_{-a}^{a}{dx\ov 2b\sqrt{1-x^2/a^2}}=\frac{a}{b} \,{\pi\ov 2}\tilde s_3^{\mathrm{(strip)}}\,.
\ee
Thus we provided a lower bound on $\tilde s_3$ which is saturated for $b/a\to 0$.


\section{Generic region in a holographic CFT$_3$}
\label{sec:generic_surface}

In this section we describe how to compute numerically the universal coefficient $\tilde s_3$ for a generic entangling region $\Sigma$ in a holographic CFT$_3$.

We consider a surface embedded in AdS$_4$ that is topologically equivalent to a disk. With reference to the coordinates (\ref{eq:AdS_coordinates}), we parametrize the surface as:
\begin{equation}\label{embedding}
 \left\{
\begin{aligned}
 &r = R(\rho, \theta)\\
 &z = Z(\rho, \theta)
\end{aligned}
\right.\,,
\quad \rho \in [0, 1]\,,\quad \theta \in [0, 2\pi]\,.
\end{equation}

Smoothness of the surface constrains the functions $R$ and $Z$ to have the following small $r$ behavior:
\es{}{
 &R(\rho, \theta) \sim \sum_\ell R_\ell\, \rho^{|\ell| + 1} e^{\ii \ell \theta}\,,\\
 &Z(\rho, \theta) \sim \sum_\ell Z_\ell\, \rho^{|\ell|} e^{\ii \ell \theta}\,.
}
The above result is most easily seen by going to Cartesian coordinates. At $\rho = 1$ we impose the boundary condition
\begin{align}
 Z(1, \theta) = 0 &&R(1, \theta) = R(\theta)\,,
\end{align}
where $R(\theta)$ defines the entangling region. For an ellipse, it is given by (\ref{eq:ellipse_R}).

The function $R$ can be chosen arbitrarily within the constraints above, and we take it to be
\es{}{
 &R(\rho, \theta) \equiv \sum_\ell R_\ell\, \rho^{|\ell| + 1} e^{\ii \ell \theta}\,,\\
  &R_\ell = \frac{1}{2\pi} \int_0^{2\pi} \dd \theta\ e^{-\ii \ell \theta} R(\theta)\,.
}

Expanding the equations of motion for $Z$ near $\rho = 1$ reveals that the $Z$ of the minimal surface is not analytic at $\rho = 1$. Instead, it can be written as
\begin{equation}\label{z asymptotics}
 Z(\rho, \theta) = \sqrt{1 - \rho^2}\, \Xi(\rho, \theta)\,,
\end{equation} 
with $\Xi$ analytic at $\rho = 1$. We represent $\Xi$ by expanding over a basis of functions:
\begin{equation}
 \Xi(\rho, \theta) = \sum_{n,\ell} \Xi_{n\ell}\ e^{i \ell \theta} \rho^{|\ell|} P_n^{(0, |\ell|)}(2 \rho^2 - 1)\,,
\end{equation} 
where $P_n^{(a, b)}$ are the Jacobi polynomials. This choice of basis is known to be good, because its elements have little linear dependence on each other, and span the space of analytic functions on the disk quite efficiently. We have to truncate the expansion at a finite value of $n$ and $\ell$, and, strictly speaking, our numerical result for $\tilde s_3$ will be a lower bound. However, if the minimal surface is sufficiently regular, the bound obtained will be very tight.

In order to find the minimal surface, it is necessary to evaluate efficiently and accurately both the area of a generic surface and its variation with respect to an infinitesimal change of the surface. This problem is complicated by the fact that the area is a divergent quantity because of the diverging conformal factor at the boundary of AdS. We now describe how this can be accomplished.

The embedding (\ref{embedding}) induces the following metric on the disk:
\begin{equation}
\begin{split}
 g_2 &= \frac{1}{Z^2}\Big[\left(R_{,\rho}^2 + Z_{,\rho}^2\right)\dd \rho^2\\
&+2\left(R_{,\rho} R_{,\theta} + Z_{,\rho}Z_{,\theta}\right)\dd \rho \dd \theta+\left(R^2 + R_{,\theta}^2 + Z_{,\theta}^2\right)\dd\theta^2\Big]\,,
\end{split}
\end{equation}
and hence the area functional is
\begin{align}
\begin{split}
 A[Z] &= \int_0^{2\pi} \dd \theta  \int_0^1\dd \rho\ \sqrt{g_2}\\
 & = \int \frac{\dd\theta \dd \rho}{Z^2}\sqrt{\left(Z_{,\theta} R_{,\rho} - Z_{,\rho} R_{,\theta}\right)^2 + R^2\left(R_{,\rho}^2 + Z_{,\rho}^2\right)}\,.
\end{split}
\end{align} 

Varying this functional with respect to $Z$ it is possible to derive a rather involved equation of motion, which is best handled with the aid of computer algebra. By expanding it about $\rho = 1$ we obtain the asymptotic behavior (\ref{z asymptotics}) for $Z$. Because of the same asymptotics we find it convenient to write
\begin{equation}
 \sqrt{g_2} = \frac{\rho}{(1-\rho^2)^{\frac{3}{2}}} a(\rho,\theta)\,,
\end{equation} 
so that $a(\rho, \theta)$ is smooth on the disk and
\begin{equation}
 a(1, \theta) = \frac{\sqrt{R^2(1,\theta) + R_{,\theta}^2(1,\theta)}}{\Xi(1,\theta)}\,.
\end{equation} 

The area integral is divergent, and needs to be regulated. The prescription is to cut off the integral at a fixed height $Z(\rho, \theta) = \delta$:
\begin{equation}
 A[Z, \delta ] = \int \dd \theta \dd \rho\ \sqrt{g_2}\ \Theta[Z(\rho,\theta) - \delta]\,,
\end{equation} 
where $\Theta$ is the unit step function.

Because the divergence is associated with the boundary of the surface, it is proportional to the perimeter $P$ of the entangling region, and we have
\begin{align}\label{regulated area}
 A[Z, \delta] \sim \frac{P}{\delta} - \tilde s_3[Z] + \mathcal{O}(\delta)\,.
\end{align} 

The divergence does not depend on $Z$ (provided the boundary conditions are satisfied), and hence does not enter the minimization procedure. In order to compute the finite quantity $\tilde s_3[Z]$ we introduce a reference integrand $a_{\text{ref}}$, which agrees with $a$ at the boundary, but whose regulated integral can be computed analytically. The \emph{finite} difference between the integral of $a$ and the integral of $a_{\text{ref}}$ can then be computed numerically with good accuracy.

We define 
\es{alDef}{
 &a_\text{ref}(\rho, \theta) \equiv \sum_\ell a_\ell\, \rho^{|\ell|} e^{i \ell \theta}\,, \\
 &a_\ell = \frac{1}{2\pi} \int_0^{2\pi} \dd \theta\ a(1,\theta) e^{-\ii \ell \theta}\,, 
}
so that $a_\text{ref}$ is a smooth function on the disk and $a_\text{ref}(1,\theta) = a(1, \theta)$. Then we write
\begin{equation}
 A[Z,\delta] = \Delta A[Z, \delta] + A_{\text{ref}}[Z, \delta]\,,
\end{equation} 
with
\es{}{
&  \Delta A[Z,\delta] = \int \frac{\dd \theta \dd \rho\,\rho}{(1-\rho^2)^{\frac{3}{2}}} \left(a - a_{\text{ref}}\right) \Theta[Z - \delta]\,,\\
& A_\text{ref}[Z,\delta] = \int \frac{\dd \theta \dd \rho\,\rho}{(1-\rho^2)^{\frac{3}{2}}}\,a_{\text{ref}}\,\Theta[Z - \delta]\,.
}
Then we have
\begin{equation}
 \Delta A[Z,\delta] = \int \frac{\dd \theta \dd \rho\,\rho}{(1-\rho^2)^{\frac{3}{2}}} (a - a_{\text{ref}}) + \mathcal O(\delta)\,,
\end{equation} 
where the integral is finite. The integral $A_{\mathrm{ref}}$ can be done analytically, and we have 
\begin{equation}
\begin{split}
 A_\text{ref}[Z, \delta] &= \frac{1}{\delta} \int_0^{2\pi} \dd \theta\ \sqrt{R^2 + R_{,\theta}^2} - 2 \pi a_{ 0} + \mathcal O(\delta)\, ,
\end{split}
\end{equation} 
where $a_0$ is given by~\eqref{alDef}.
Collecting all the pieces we have
\begin{align}
\begin{split}
  A[Z,\delta]&=\frac{1}{\delta} \int_0^{2\pi} \dd \theta\ \sqrt{R^2 + R_{,\theta}^2}+
 \int  \frac{\dd \theta \dd \rho\,\rho}{(1-\rho^2)^{\frac{3}{2}}}\left[a - a_{\text{ref}}\right] \\&- \int_0^{2\pi} \dd \theta\ a(1,\theta) + \mathcal{O}(\delta)=\frac{P}{\delta} - \tilde s_3[Z] + \mathcal{O}(\delta)\,,
\end{split}
\end{align} 
where we recognized in
\begin{equation}
 P \equiv \int_0^{2\pi} \dd \theta\ \sqrt{R^2 + R_{,\theta}^2}
\end{equation} 
the perimeter of the entangling region. Using computer algebra, it is possible to compute explicitly also the variation of $A$ with respect to a change in $Z$, without forgetting that both $a$ and $a_\text{ref}$ depend on $Z$. 

The integrals are conveniently evaluated using Gaussian quadrature. In particular, the double integrals have the form
\begin{equation}
I = \int_0^{2\pi} \dd \theta \int_0^1 \dd \rho\  \frac{\rho}{(1-\rho^2)^{\frac{1}{2}}} f(\rho, \theta)\,,
\end{equation}
with $f$ smooth on the disk. We evaluate them as
\begin{equation}
I = \frac{\pi}{n}\sum_{i = 1}^{m}\sum_{j=1}^{n} w_i f(\rho_i, \theta_j)\,,
\end{equation} 
where $\theta_j = 2 \pi j / n$ and $w_i$, $\rho_i$ are the weights and collocation points of the Gaussian quadrature associated with the  measure
\begin{equation}
 \int_{-1}^1 \dd \rho\  \frac{|\rho|}{(1-\rho^2)^{\frac{1}{2}}}\,.
\end{equation} 

We have now cast the problem to the numerical minimization of a real function of many variables (the coefficients $\Xi_{n\ell}$), of which we know explicitly the gradient. Therefore, we search for a minimum using the conjugate gradient method.

\section*{Acknowledgments}
We thank  A.~Lewkowycz, E.~Dyer, J.~Lee, E.~Perlmutter, S.~Pufu, and V.~Rosenhaus, and especially H.~Liu for useful discussions. We thank A.~Lewkowycz and E.~Perlmutter for proposing the generalization to R\'enyi entropies and sharing~\cite{aitor} with us before publication. AA was suppported by DE-FG02-05ER41360, DE-FG03-97ER40546, by the Alfred P. Sloan Foundation, and by the Templeton Foundation. MM was supported by the U.S. Department of Energy under cooperative research agreement Contract Number DE-FG02-05ER41360\@.  Simulations were done on the MIT LNS Tier 2 cluster, using the Armadillo C++ library.

\bibliographystyle{ssg}
\bibliography{shape}

\end{document}